\documentclass[aps,prb,two column,superscriptaddress]{revtex4-1}
\usepackage{gensymb}
\usepackage{hyperref}
\usepackage{graphicx}
\usepackage{amsmath}
\usepackage{color}
\usepackage{ulem}

\begin{document}
\newcommand{\Ss}{\textsubscript}
\newcommand{\Us}{\textsuperscript}
\newcommand{\STO}{SrTiO\Ss{3}}
\newcommand{\GTO}{GdTiO\Ss{3}}
\newcommand{\LAO}{LaAlO\Ss{3}}
\newcommand{\LTO}{LaTiO\Ss{3}}
\newcommand{\RTO}{ReTiO\Ss{3}}
\newcommand{\NGO}{NdGaO\Ss{3}}
\newcommand{\LCO}{LaCrO\Ss{3}}
\newcommand{\PyroCl}{Re\Ss{2}Ti\Ss{2}O\Ss{7}}
\newcommand{\ETO}{EuTiO\Ss{3}}
\newcommand{\ox}{O\Ss{2}}
\newcommand{\tit}{Ti\Us{3+}}
\newcommand{\tif}{Ti\Us{4+}}
\newcommand{\DC}{\degree C}
\newcommand{\mx}{\times}
\newcommand{\rev}{\cite}
\newcommand{\RAN}{$R^{xy}_{AN}$}
\newcommand{\dxy}{3d\Ss{xy}}
\newcommand{\dxyz}{3d\Ss{xz/yz}}
\newcommand{\dxyzo}{d\Ss{xz/yz}}
\newcommand{\dxyo}{d\Ss{xy}}
%\preprint{}

\title{Inhomogeneous superconductivity and quasilinear magnetoresistance at amorphous \LTO/\STO{} interfaces}


\author{N. Lebedev}
 \email[Corresponding author:]{lebedev@physics.leidenuniv.nl}
%\thanks{}
\affiliation{Huygens - Kamerlingh Onnes Laboratory, Leiden University, P.O. Box 9504, 2300 RA Leiden, The Netherlands}
\author{M. Stehno}
\affiliation{Physikalisches Institut (EP 3), Universität Würzburg, Am Hubland 97074 Würzburg, Germany}
\author{A. Rana}
\affiliation{School of Engineering and Technology, BML Munjal University (Hero Group), Gurgaon, India - 122413}
\author{N. Gauquelin}
\author{J. Verbeeck}
\affiliation{Electron Microscopy for Materials Science, University of Antwerp, Campus Groenenborger Groenenborgerlaan 171, 2020 Antwerpen, Belgium}
\author{A. Brinkman}
\affiliation{MESA+ Institute for Nanotechnology, University of Twente, P.O. Box 217, 7500 AE Enschede, The Netherlands}
\author{J. Aarts}
\affiliation{Kamerlingh Onnes Laboratory, Leiden University, P.O. Box 9504, 2300 RA Leiden, The Netherlands}

\date{\today}

\begin{abstract}
We have studied the transport properties of \LTO/\STO{} (LTO/STO) heterostructures. In spite of 2D growth observed in reflection high energy electron diffraction, Transmission Electron Microscopy images revealed that the samples tend to amorphize. Still, we observe that the structures are conducting, and some of them exhibit high conductance and/or superconductivity. We established that conductivity arises mainly on the STO side of the interface, and shows all the signs of the 2-dimensional electron gas usually observed at interfaces between STO and LTO or \LAO{}, including the presence of two electron bands and tunability with a gate voltage. Analysis of magnetoresistance (MR) and superconductivity indicates presence of a spatial fluctuations of the electronic properties in our samples. That can explain the observed quasilinear out-of-plane MR, as well as various features of the in-plane MR and the observed superconductivity.
\end{abstract}

%\pacs{}

\maketitle

\section{Introduction}
%\subsection{}
%\subsubsection{}
Since the discovery of conductivity\rev{OhtomoNAT} at the interface between the two nonmagnetic band insulators \LAO{} (LAO) and \STO{} (STO), oxide interfaces have been under intense investigation.
The dominant view in the literature on the origin of conductivity at the (001) LAO/STO interface is the so-called polar catastrophe scenario\rev{NakagawaNMAT, GariglioAPL}, based on the difference between the stacking of neutral layers in STO, but 1-electron-charged layers in LAO. To avoid the discontinuity at the interface, half an electron per unit cell has to transfer from the LAO surface down to interface, leading to a formation of two-dimensional electron liquid (2DEL). Besides that, also La/Sr intermixing\rev{WillmottPRL} and oxygen vacancies formed in the STO\rev{HerranzPRL, KalabukhovPRB} can lead to the creation of the conducting layer. Moreover, it was proposed recently that the development of a critical density of oxygen vacancies at the surface of the LAO layer plays a vital role in avoiding polar discontinuity\rev{GariglioAPL, YuNCOM}.

Along with LAO/STO, also the interface between the antiferromagnetic Mott insulator \LTO{} (LTO) and STO has been under intensive investigation. LTO is polar along (001) crystal direction, so a charge transfer similar to LAO/STO may be expected. At the LTO/STO interface, the polar discontinuity can be resolved by the variable valence of Ti\rev{OhtsukaAPL, YouPRB}. Indeed, Biscaras et al. [\onlinecite{BiscarasNCOMM}] argued that conductance at this interface is on the STO side, similar to LAO/STO. On the other hand, Wong et al. [\onlinecite{WongPRB}] proposed that the LTO layer is metallic when grown on STO, due to a lattice distortion induced by stress.
La/Sr intermixing\rev{VilquinApSS,TokuraPRL,Goodenough,HaysPRB}, and oxygen and lanthanum off-stoichiometry\rev{GariglioPRB} can also lead to conductivity in LTO. Furthermore, a recent study by Scheiderer et al. [\onlinecite{ScheidererAM}] has shown that the LTO layer in LTO/STO heterostructures is suffering from strong overoxidation due to a migration of oxygen from STO and oxidation in the~air in uncapped films. Such processes are able to transform the LTO layer into an amorphous state. The amorphous oxide interfaces were shown to be conducting due to oxygen vacancies formed on the surface of STO \rev{LeeNL, LiuPRX, ChenNL}, and, similar to the stoichiometric crystalline interfaces\rev{ReyrenScience, CavigliaNature}, the amorphous interfaces are also superconducting\rev{FuchsAPL, PrawiroatmodjoPRB}.

In this paper, we have studied LTO/STO interfaces grown by Pulsed Laser Deposition (PLD), and found that in spite of layer-by-layer growth signatures, the LTO layer tends to amorphize. Still, the conductivity in the system is basically due to a 2DEL formed on the STO side of the interface. The 2DEL properties are not much different from those of other STO-based oxide interfaces. In particular, Hall data show two-band behavior with standard values for the carrier concentrations and back-gating shows the presence of a Lifshitz point. Less normal is a quasilinear Magnetoresistance (MR), and non-uniform superconductivity. We argue that the possible origin of these phenomena is the non-uniform distribution of oxygen vacancies on the STO surface due to the uncontrolled oxidation process in the LTO layer, which lead to spatial inhomogeneities. This inhomogeneity is clearly seen in the superconducting state, but not easily discernible in the normal state, which is an important part of the message.

\section{Experimental details}

\begin{figure*}
\includegraphics[scale=0.4]{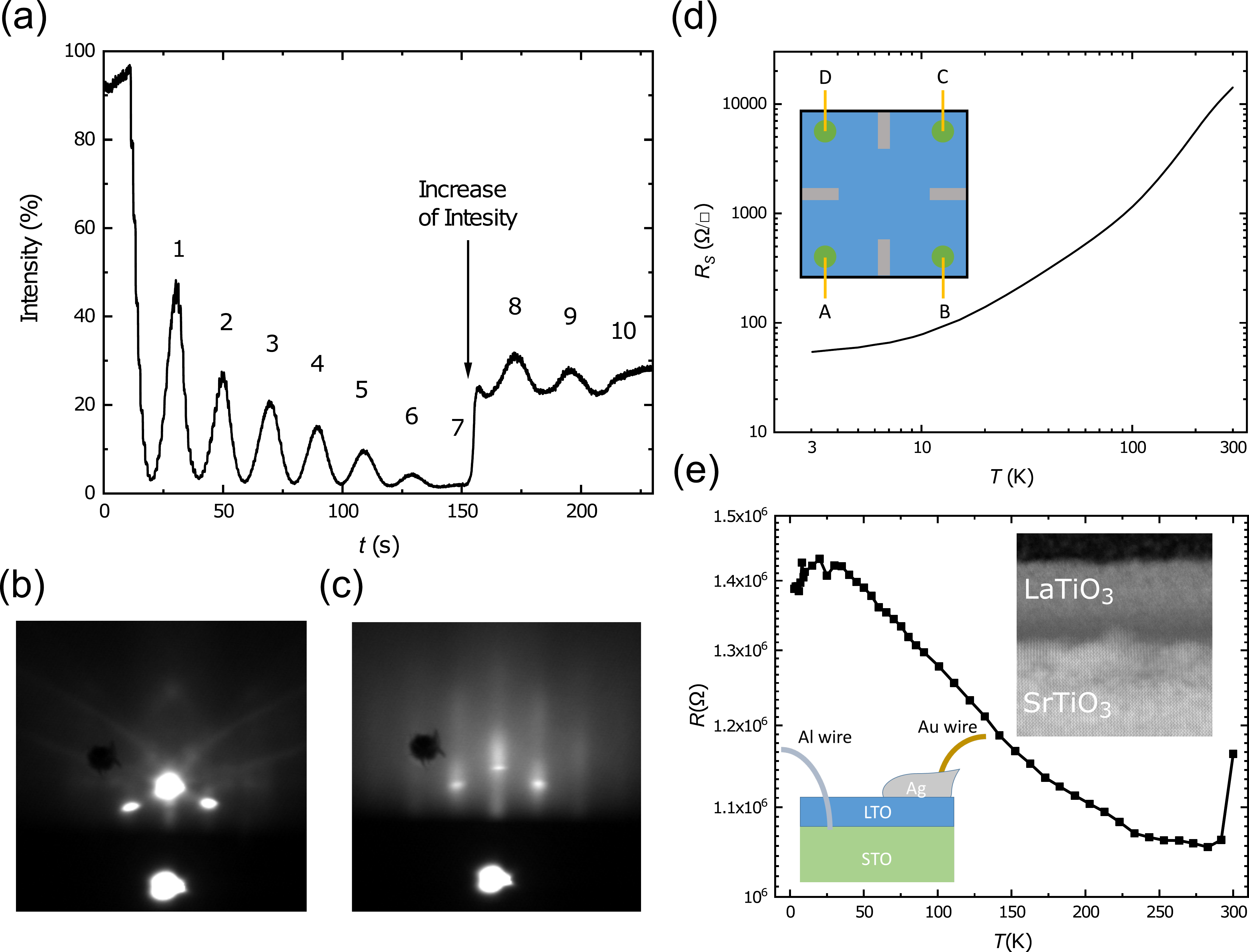}
\caption{(a) RHEED intensity monitoring during grown 10 u.c. of LTO. RHEED patterns (b) before and (c) after deposition. (d) Temperature dependence of Sheet resistance. Insert: Sketch of van der Pauw measurements. (e) Temperature dependence of the two-probe measured resistance of the~LTO layer. Schematics of two-probe measurement and STEM scan of the~LTO/STO interface are shown on the inserts.\label{Fig1}}
\end{figure*}

\begin{figure*}[t]
\includegraphics[scale=0.26]{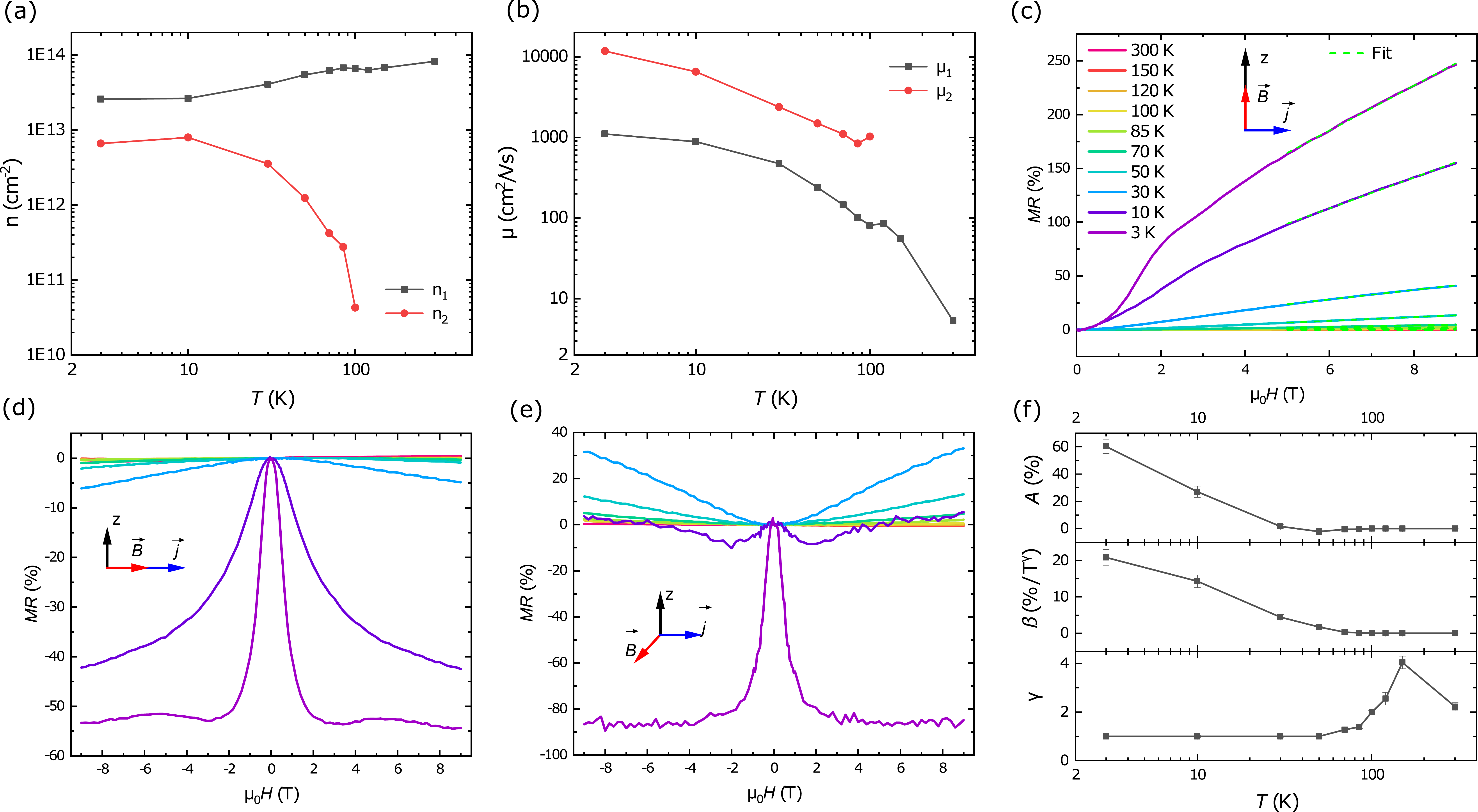}
\caption{(a) Carrier concentration and (b) mobility versus temperature obtained from a two-band analysis of Hall resistance measurements (given in the Supplement). (c-e) The magnetoresistance MR with magnetic field oriented (c) perpendicular to the sample plane, (d) in-plane parallel to the current direction, and (e) in-plane perpendicular to the current direction. (f) Temperature dependence of the parameters to describe the non-linear out-of-plane MR by fitting Eq.~\ref{eq1}.  \label{Fig2}}
\end{figure*}

LAO layers were grown by Pulsed Laser Deposition on a TiO$_2$-terminated surface of STO(001) single crystal substrates. The growth temperature was 750 \DC{}. Growth was in an \ox{} atmosphere utilizing two nominal pressures: $1\mx {10}^{-4}$ and $5\mx {10}^{-4}$ mbar. The thickness of the samples was determined by observing the intensity oscillations using Reflection High Energy Electron Diffraction (RHEED) and fixed at 10 u.c. (see Fig. \ref{Fig1}a).
The RHEED pattern showed characteristic stripes indicating 2D growth (Fig. \ref{Fig1}b,c). Magnetotransport measurements in the range 3-300 K were performed with a physical properties measurement system (a  PPMS) from Quantum Design, and below 1 K in an Oxford Instruments Triton dilution refrigerator.
Samples were wirebonded with Al wire for magnetotransport measurements, and measured with a standard lock-in technique. Scratches were made on the samples by a diamond knife in the center of each edge to ensure the current path through the sample center, as shown schematically in the inset Fig. \ref{Fig1}d, together with a denumeration of the contacts.

Most of the measurements were performed in the van der Pauw (VDP) geometry. To determine the sheet resistance, two resistances were measured, one called R$_H$ with the current applied over one edge (contacts A,B) and the voltage measured along the opposite edge (contacts C,D), and one called R$_V$ using the other pair of edges (current through A,D, voltage over B,C). The sheet resistance R$_S$ was then calculated by by solving the VDP equation for R$_S$ by the Newton-Raphson method :

\begin{equation}
    e^{-\pi R_V / R_S} + e^{-\pi R_H / R_S} = 1
\end{equation}

The magnetoresistance was determined in the same way, by either applying in-plane or out-of-plane fields. Hall data were obtained by injecting the current along one diagonal and measure the voltage across the other one, using an out-of-plane field. The out-of-plane magnetotransport data were (anti-)symmetrized. The in-plane data were not. Instead, the two measured voltages in in-plane geometry were used to obtain MR with the current parallel and perpendicular to the current direction. The experimental data obtained at temperatures below 1 K were smoothed to remove noise except for the measurements in magnetic field. The geometry for the measurements of the superconducting transition in the Triton is described in Section \ref{SC}. An extra sample was prepared for study by scanning transmission electron microscope (STEM), using an oxygen pressure of $5\mx {10}^{-4}$ mbar.
The conductivity of the LTO layer was checked by using additional gold wires, which were glued by silver paint to the surface of the sample, and resistance was measured by a source meter with an applied current of $1 \mu A$ in a two-probe geometry.

\section{Normal state magnetotransport}
\subsection{The origin of conductance\label{Cond}}
The different samples did shown a variation in conducting properties. Some exhibit higher conductance and/or superconductivity. We did not observe a correlation between high conductance or superconductivity and the oxygen pressure during growth. The transport data reported here is on a sample which shows high conductivity, a decrease of the sheet resistance upon lowering the temperature (Fig. \ref{Fig1}d) with a large residual resistivity ratio RRR $ = R_S(300 K)/R_S(10K)=261$, and superconductivity below 300 mK.

As mentioned above, the conductance in these heterostructures can arise not only from a 2DEL forming at the STO/LTO interface but also in the LTO itself. To distinguish between these two possibilities, after performing the transport measurements presented below, we investigated the conductivity of the LTO layer in the following manner. A Au wire was glued by the silver paint to the LTO surface as is shown schematically in the inset in Fig. \ref{Fig1}e. Resistance measurements as function of temperature between the Al wire contact and the Au wire contact, shown in Fig. \ref{Fig1}e, demonstrated that although the LTO layer is slightly conducting, it exhibits insulating behavior going to lower temperatures. That conductance could arise due to the formation of pinholes in the LTO film under the surface of silver paint\rev{KumarAPL}. Moreover, results of Scanning Transmission Electron Microscopy (STEM) (see inset in Fig. \ref{Fig1}e) reveal that the LTO layer in our samples is amorphous, in agreement with the~results of Scheiderer et al. [\onlinecite{ScheidererAM}]. Because the LTO layer is (almost) insulating and amorphous, we conclude that the conductivity in our samples arise from oxygen vacancies on the~surface of STO similar to the previously reported conducting interfaces between amorphous oxide and STO\rev{LeeNL, LiuPRX, ChenNL}.
This can explain the high $RRR$ but also the variation of conducting properties observed from sample to sample.

\subsection{Magnetotransport without back gate\label{MR}}
\begin{figure}[b]
\includegraphics[scale=0.26]{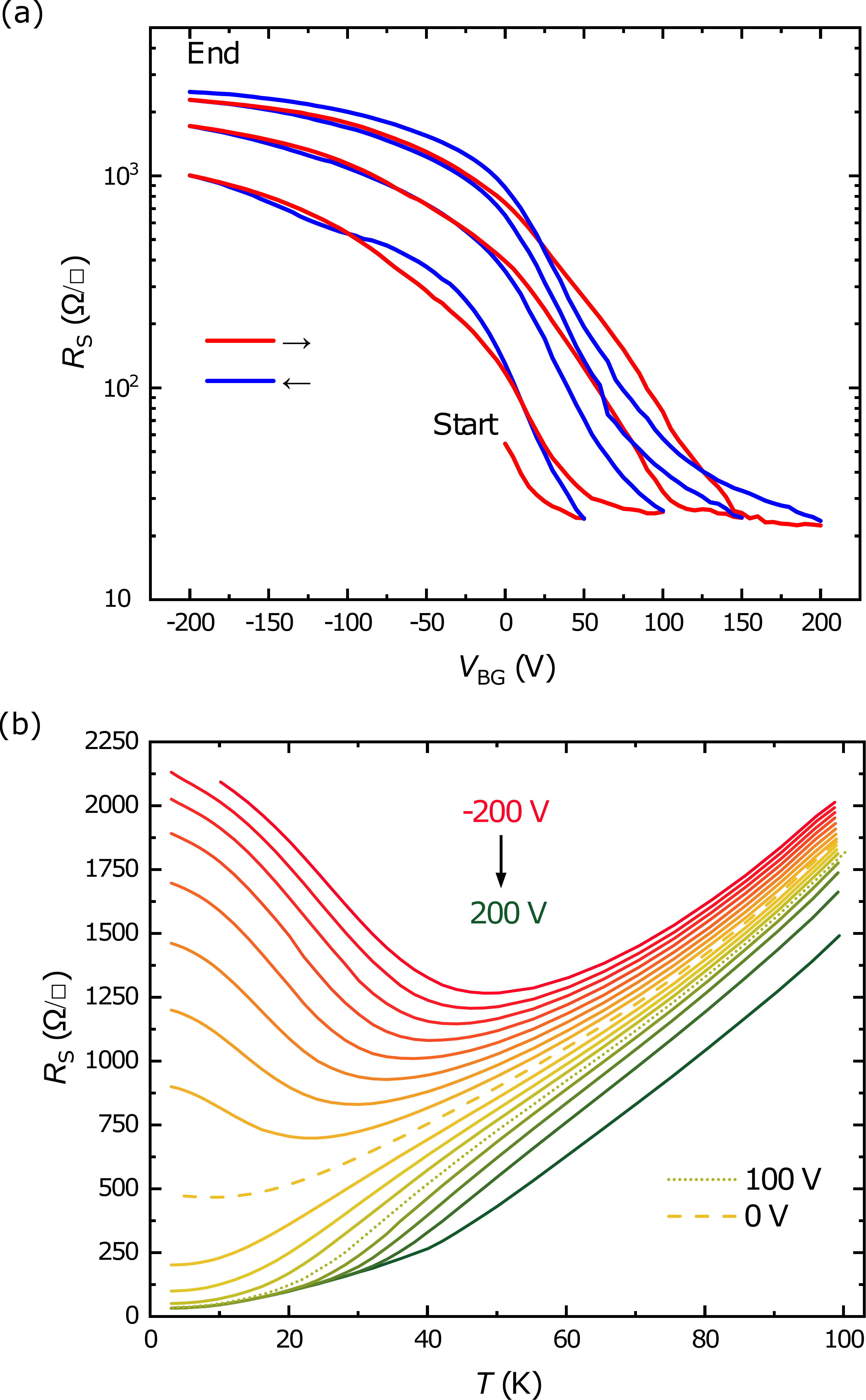}
\caption{(a) Dependence of the sheet resistance $R_S$ on the back gate voltage. (b) Temperature dependence of $R_S$ for gate voltages from $200$ to $-200$ V with steps of $25$ V. \label{Fig3}}
\end{figure}
Broadly speaking, the magnetotransport properties are similar to previously reported results on oxide heterostructures. In particular the Hall resistance becomes non-linear below 100 K, marking the appearance of two-band behavior, with two types of carriers: of high concentration and low mobility, and vice versa. The Hall data and details of the Hall analysis are given in the Supplement, extracted carrier concentrations and mobilities in Fig.~\ref{Fig2}a-b.
The out of plane MR is anomalous. It is almost flat at high temperatures, and in low fields gradually becomes parabolic with lower temperature. So far, such behaviour is similar to most of the results on STO-based interfaces. However, below 70 K, a quasilinear MR in high fields starts to develop (Fig. \ref{Fig2}c), with values much higher than reported previously in LTO/STO\rev{DasPRB}. To describe this behavior, we fitted the MR in the field range form 5~T to 9~T with the following equation:
\begin{equation}\label{eq1}
MR= A+\beta B^{\gamma},
\end{equation}
where $A, \beta, \gamma$ are fitting parameters. The results of the fit are shown in Fig.~\ref{Fig2}f. At high temperature where the MR is small, the parameters $A$ and $\beta$ are almost zero. At low temperatures, $\gamma$ is smaller than $2$, indicating that linear contribution to MR becomes dominant. Note that for this analysis, we limited the lowest boundary for $\gamma$ to 1 in order to avoid unphysical behaviour of $A$.

The in-plane MR is negligible at high temperatures (Fig. \ref{Fig2}d,e). At low temperatures, the parallel-to-current configuration shows a negative MR, which increases at temperatures below 30 K and undergoes a transition from parabolic to bell shape. The perpendicular-to-current configuration exhibits first an increase of the positive MR down to 70 K, shows the onset of negatives lobes below 30 K and finally transforms also to a bell shape with saturation at 3 K. Note that the VDP configuration does not allow to reliably exclude contributions to the MR of currents perpendicular to the magnetic field in the parallel in-plane geometry and currents parallel to the field in the perpendicular in-plane geometry.

\begin{figure*}
\includegraphics[scale=0.26]{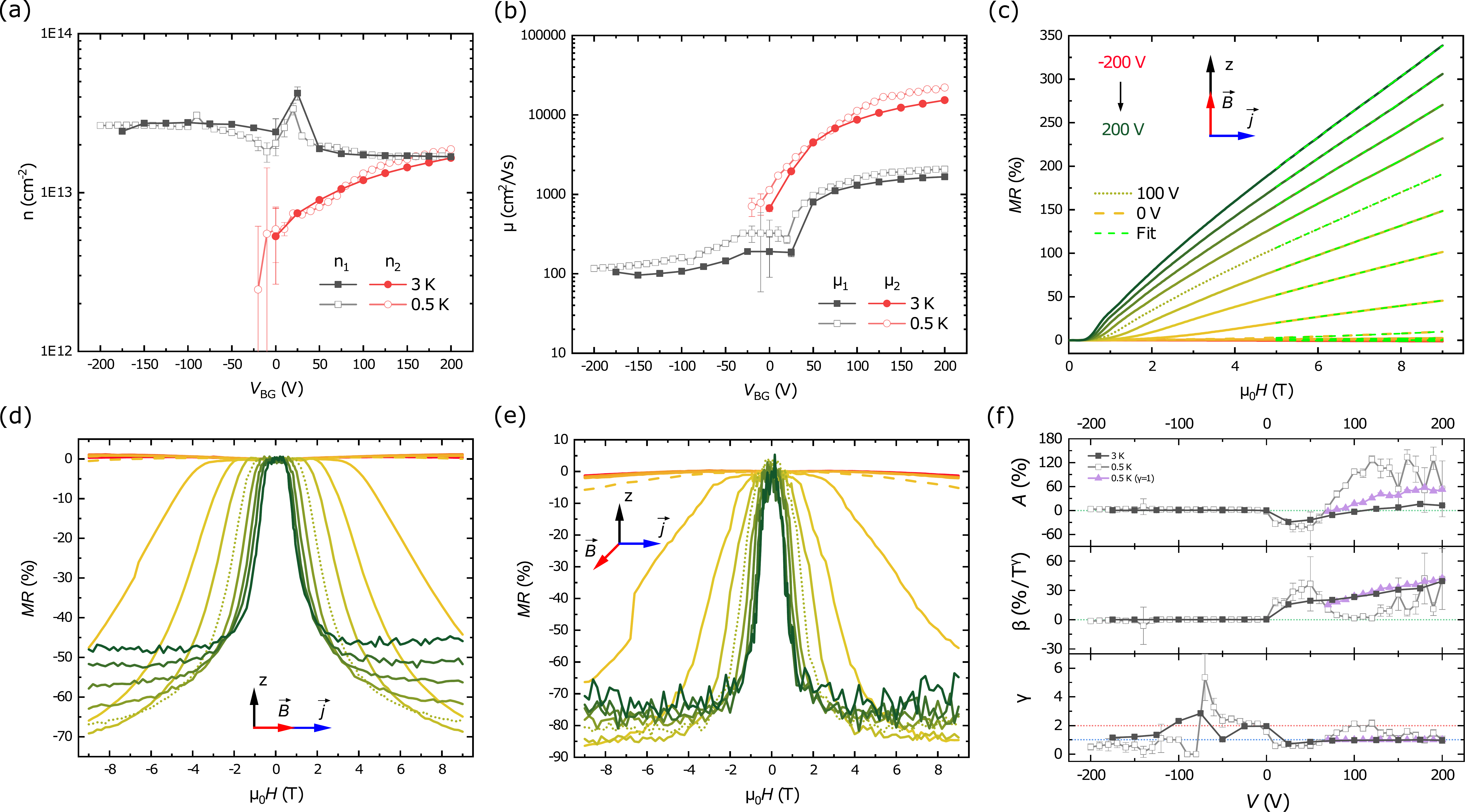}
\caption{(a) Carrier concentration and (b) mobility versus gate voltage at the temperatures 3~K and 0.5~K as indicated,  obtained from a two-band analysis of Hall resistance measurements. (c-e) The magnetoresistance MR at 3~K with magnetic field oriented (c) perpendicular to the sample plane, (d) in-plane parallel to the current direction, and (e) in-plane perpendicular to the current direction. Note that $-200$ V is not shown for out-of-plane magnetic field measurements. (f) Temperature dependence of the parameters to describe the non-linear out-of-plane MR by fitting Eq.~\ref{eq1}. Note that near transition from negative to positive high field MR in the~range $[-110,-80]$ parameter $\gamma$ and, for some values, $\beta$ were fixed. \label{Fig4}}
\end{figure*}

\subsection{The effect of gating on the sheet resistance\label{G-sheet}}
To further study the magnetotransport properties, we applied a back gate voltage V$_{BG}$ to the sample. First we investigate the effect of a gate voltage on R$_S$. The "training" of the sample at 3 K, meaning successive up-down sweeps of the voltage, (Fig.~\ref{Fig3}a) showed an increase of R$_S$ in the backsweeps, which is usually explained as the trapping of charges escaping from the quantum well \cite{BiscarasSREP,ChunhaiPRL}. We observe some hysteric effects between the up sweep and the subsequent down sweep which are not always present; moreover, we do not find the interface to become insulating in the backsweep at low or negative V$_{BG}$. This was found for highly conducting (crystalline) interfaces \cite{CavigliaNature,LiaoPRB}, but not for less conducting ones \cite{BiscarasSREP,ChunhaiPRL}. Fig.~\ref{Fig3}b shows the temperature dependence of R$_S$, measured from 200~V down to -200~V. Coming from negative V$_{BG}$, the R$_S$ shows an upturn to low temperatures which disappears at 0~V. Also, the change in R$_S$ at low temperatures is largest between 0~V and 100~V, similar to what is seen in the training sweeps shown in Fig.~\ref{Fig3}a. We will come back to this behavior in the discussion.

\subsection{The effect of gating on the magnetotransport\label{G-MR}}

Starting again with the Hall resistance, we find it becomes nonlinear between $-25$ and $0$ V (Supplement Fig. S2b),  signaling the well-known Lifshitz transition \rev{JoshuaNCOM, SanderPRL}. The gate dependence of the carrier concentrations and mobilities, found after standard analysis, is given in Fig~\ref{Fig4}a,b. In the proximity of the transition,  between $-20$ and $40$ V, the two-band model gives an anomalous increase of carrier concentration and a dip in the mobility of the majority carriers, with high error bars. This is the case at 3~K, as well as at 0.5~K, with the measurements performed in a different cryostat. This anomaly probably arises due to a fast decrease in the second type of  carriers, which the fit is not able to correctly describe; and to the fact that the mobility values in this regime are close, which complicates the fitting procedure. To avoid such problems, we limited the lowest possible mobility value of majority carriers in this region by the value extracted from one band analysis at the closest point to the transition. Such a limit resulted in a plateau of the mobility of majority carriers versus V$_{BG}$ near the Lifshitz transition. Note also that the carrier concentrations of the two bands become almost equal above 100~V.
%
%Although this approach provides more consistent results than without the limit, some increase of carrier concentration of majority carriers still can be seen in Fig. \ref{Fig5}h.
%Above 100 V, mobilities of both carriers, but especially the high mobile carriers, are significantly enhanced.
%Moreover, above 70 V, the carrier concentration values became very close to each other and even crossed in this region at 0.5 %K.

Turning to the MR at 3~K, the out-of-plane MR, shown in Fig~\ref{Fig4}c (See Supplement Fig. S2a for a zoom-in around low fields and MR values), is small and negative in high fields at high negative gate voltages. In this range of V$_{BG}$, the parameters $A$ and $\beta$ are almost zero(Fig. \ref{Fig4}f), and Eq. \ref{eq1} is not always adequate to describe the high field  MR; also $\gamma$ shows inconsistent behavior.
However, with an increase of the gate voltage, MR becomes positive, and above 50~V, the quasilinear MR at 3 K (Fig. \ref{Fig4}c) starts to develop with the value of $\gamma$ about $1$ (Fig. \ref{Fig4}f). At 0.5 K, Eq. \ref{eq1} gives poorer fit with higher error bars and less clear gate dependence. That can be due to more noise in the data obtained in our low temperature cryostat due to low current used and smaller available field range ($[-8 T, 8 T]$). However, if we fix $\gamma=1$ starting from $70$ V, then the fit results are consistent (purple curves in Fig. \ref{Fig3}d).
Linear high field MR has been seen before in STO-based heterostructures\rev{BenShalomPRL, JoshuaNCOM, FleskerPRB}.\\
The in-plane MR parallel to the current shows a transition from positive to negative at $0$ V, whereas the in-plane MR perpendicular to the current stays negative (Fig~\ref{Fig4}d,e and Supplement Fig. S2c,d). Above $0$ V, both in-plane configurations showed substantial enhancement of the negative MR and developed the bell shape field dependence (Fig. \ref{Fig4}d,e). They exhibit saturation in high fields above 100~V, and the amplitude starts to decrease, especially in the configuration field parallel to the current. \\
Summarizing this part, the normal state properties show all the characteristics of the oxide 2DEL, with a high conductance due to a high carrier concentration, and a Lifshitz point around zero gate voltage. The MR is clearly sensitive to the Lifshitz point and in particular in the out-of-plane configuration shows quasi-linear behavior which needs to be discussed.

\section{Electronic transport in the superconducting state \label{SC}}
\begin{figure}
\includegraphics[scale=0.18]{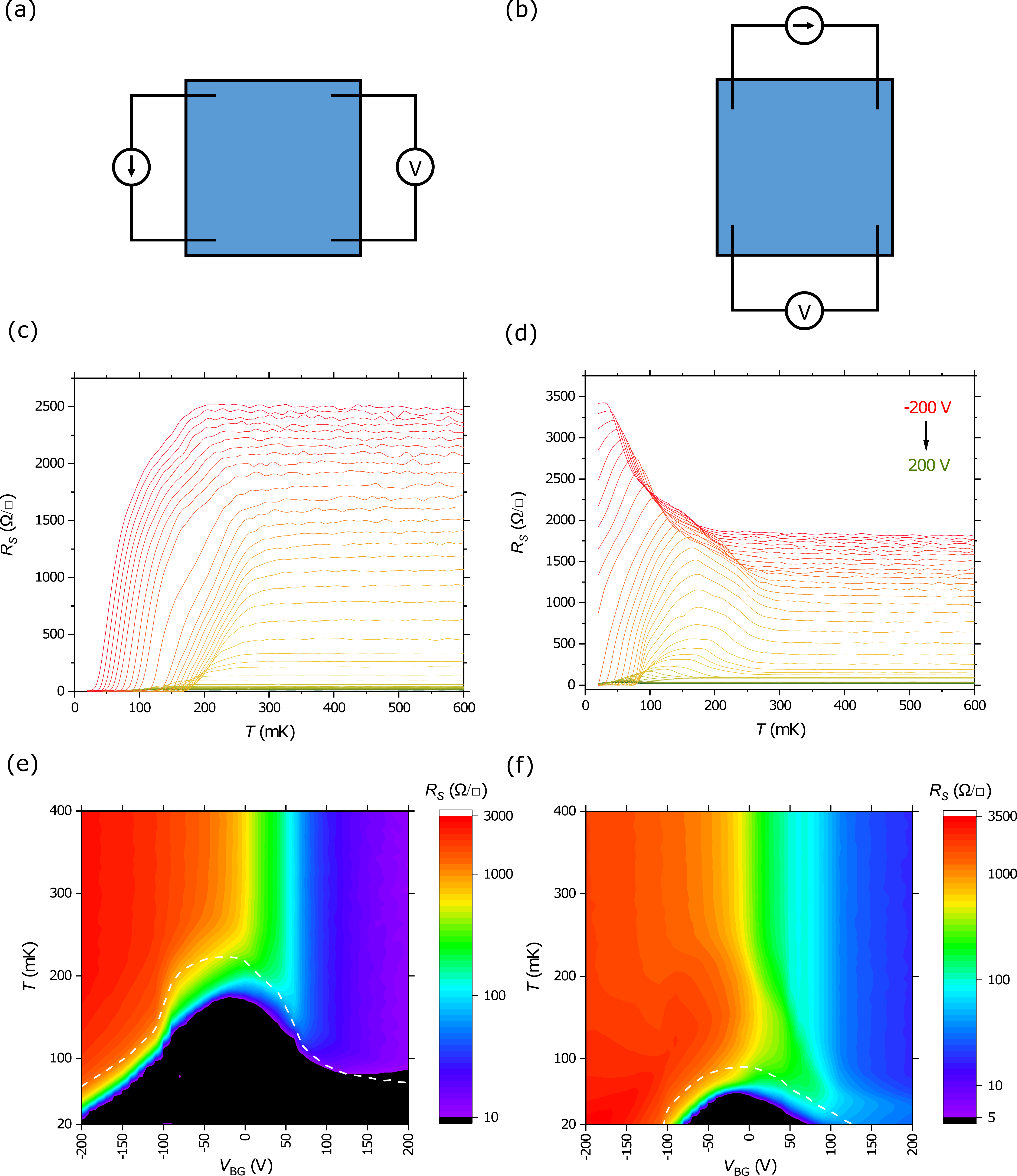}
\caption{Behavior of the sheet resistance R$_S$ as function of temperature T and back gate voltage V$_{BG}$ for two van der Pauw contact geometries called (a) 'vertical' and (b) 'horizontal'. $R(T)$ curves at different gate voltage with step of $10$ V for (c) vertical and (d) horizontal configurations. The same data visualized in colour map form for (e) vertical and (f) horizontal configurations.\label{Fig5}}
\end{figure}

\begin{figure*}
\includegraphics[scale=0.33]{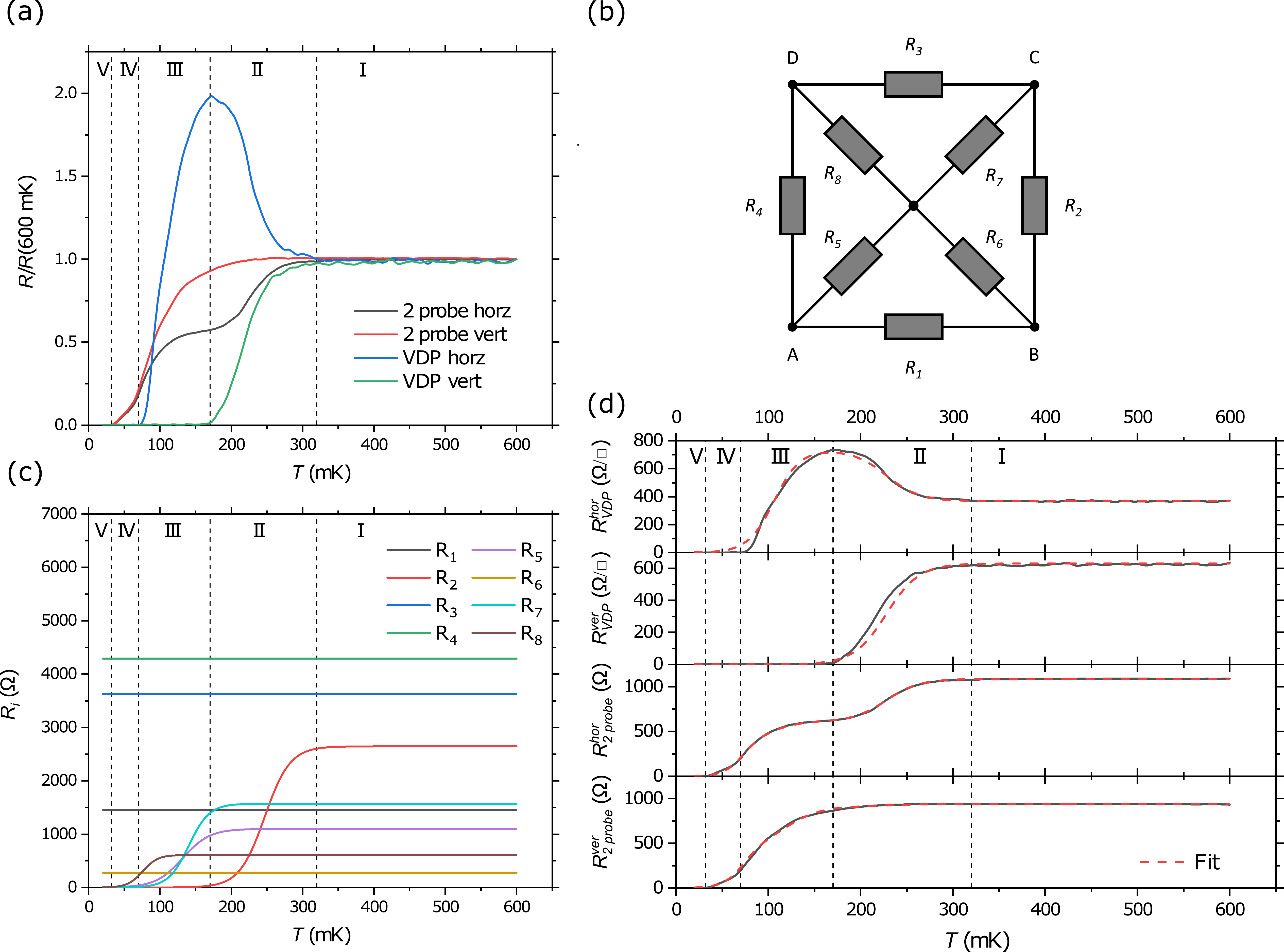}
\caption{(a) Normalized resistance for the different measured configurations at 0 V. (b) Sketch of schematics used to the model behaviour of the system. (c) Temperature dependence of Ri. (d) Resistance in different measured configurations and fit using Eq. 5-8. \label{Fig6}}
\end{figure*}

We studied the superconducting properties of the sample in the VDP geometry, using either the 'horizontal' or the 'vertical' sides, and for the whole range of gate voltages V$_{BG}$. We also measured in a two-probe configuration (current and voltage contacts on the same side). Those data are given in the Supplement, Fig.S3. We find dissimilar behavior in the two VDP measurements, so we did not calculate a sheet resistance R$_S$ by solving the VDP equation. Instead, we multiplied the measured resistance by the VDP constant c$_{VDP} = \frac{\pi}{\ln2}$. In Fig.~\ref{Fig5}, we represent the data in two different ways. Fig.~\ref{Fig5}c,d show R$_S(T)$ for gate voltage between -200~V and 200~V. Fig.~\ref{Fig5}e,f shows R$_S$ in a colorscale, as function of V$_{BG}$ and T. In the vertical configuration, the resistive transition is more or less monotonous, as can be expected. T$_c$ increases when V$_{BG}$ is increased from -200~V, reaches a maximum around 0~V, and then decrease again. At the same time, R$_S$ decreases continuously. The behavior of T$_c$ at high V$_{BG}$ can therefore be better followed in the colorscale plot, where it is shown as a dashed line marking a 50\% drop from the resistance at 600~mK. In the horizontal configuration, the resistance around $T_c$ is non-monotonous. For all V$_{BG}$, the resistance first rises before going down to 0. Comparing the color plots, both measurements show a dome shaped T$_c$ behavior similar to reported previously \cite{CavigliaNature, JoshuaNCOM}, with a maximum around 0~V, but the maximum T$_c$ is much lower in the horizontal configuration. \\

Anisotropy in STO-based structures has been reported before\rev{BiscarasPRL,FuchsAPL}. It can arise, for instance, due to the formation of regions with different conducting properties, which strongly affects measurements in the VDP geometry. In a recent report on the effect of STO domain walls on the normal state resistance of mesoscopic LAO/STO devices, the authors of [\onlinecite{GobleSREP}] proposed a scenario where a high resistance region develops in the center of the sample in order to explain the anisotropic behavior they observed. In our case, the behaviour of $R(T)$ dependencies above 0.3 K does not differ significantly for both geometries, although some variation of the resistance is present. In the transition, however, the sample may well become inhomogeneous. The two-probe resistance behavior in the Supplement shows indications of a percolative transition, and features we observe can be understood using a resistor model for an inhomogeneous superconductor adapted from Ref. [\onlinecite{VaglioPRB}]. The original model was precisely used to explain the peak in R$_S$(T) for films measured in the VDP geometry\rev{VaglioPRB}. A sketch of the equivalent electric circuit for the modified model, where all resistances have different transition temperatures, is shown in Fig. \ref{Fig6}b. The sample corners in Fig. \ref{Fig4}a-d are designated as in the insert of Fig. \ref{Fig1}e. The algorithm to solve the equations is described in the Supplement. \\

The normalized resistances at 0~V for the different measurement configurations, including the 2-probe measurements, are plotted in Fig. \ref{Fig6}a. They can be divided into five regions. In region $I$, the temperature is above $T_{c}$ for all percolation paths, and all resistances are in the normal state. $R_{VDP}^{ver}$ decreases in region $II$ and becomes zero in region $III$, while $R_{VDP}^{hor}$ reduces to zero in region $III$. In region $IV$, both two probe resistances become equal to each other and reach zero at the start of region $V$. Of course, multiple combinations of transition temperatures of $R_i$ can yield this behavior. The temperature dependencies of $R_i$ that lead to a very good fit of the data are shown in Fig.~\ref{Fig6}c. The fits themselves are shown in Fig.~\ref{Fig6}d. The table with fit parameters is included in the Supplement.

In region $II$, $R_{2}$ goes to zero and, therefore, $R_{VDP}^{ver}$ goes to zero too. Also, the denominator decreases faster than the numerator in Eq. S6 and consequently, $R_{VDP}^{hor}$ now increases. The opposite trend is observed for $R_{2 probe}^{hor}$, whereas $R_{2 probe}^{ver}$ changes insignificantly. In region $III$, $R_5$ and $R_7$ reduce to zero, and thereby $R_{VDP}^{hor}$ reduces to zero. In region $IV$, $R_{2}=0$, $R_{5}=0$ and $R_{7}=0$, as well as the measured resistances $R_{VDP}^{hor}$ and $R_{VDP}^{ver}$, and therefore $R_{2 probe}^{hor}$ and $R_{2 probe}^{ver}$ become equal:
\begin{equation}\label{eq2}
R_{2 probe}^{hor}=R_{2 probe}^{ver}=\frac{R_{1}R_{3}R_{4}R_{6}R_{8}}{R_{1}R_{6}(R_{3}R_{4}+R_{3}R_{8}+R_{4}R_{8})},\\
\end{equation}
The resistances $R_1$ and $R_6$, occurring as a product in both numerator and denominator, have to remain finite in the measured range, for eq.\ref{eq2} to be determinate. In region $V$, one of resistances $R_{3}$, $R_{4}$ or $R_{8}$ is zero because the resistances in the two-probe are zero.
In our case, it is $R_8$, whereas $R_1$, $R_3$, $R_4$, and $R_6$ are assumed not to undergo a superconducting transition in the measured range of temperatures to stabilize the fit. \\

The behavior on both sides of the resistance dome around zero gate voltage, for our different measurement configurations, can be understood from this model, assuming the $T_c$'s of all percolation paths on both sides of the dome are suppressed by the gate voltage. For the VDP vertical configuration, because $R_{2}$ has the higher $T_c$, the resistance stays zero in the whole range of gate voltages. In the other configurations, since $R_5$, $R_7$ and $R_8$ stay finite, $T_c$ is (more) quickly suppressed, both in the VDP horizontal and in the two-probe configurations.

The proposed model also provides insight into the large critical currents observed in our sample, shown in Supplement Fig. S4f,g. The percolation paths for critical currents corresponding to $R_{2}$, $R_{5}$ and $R_{7}$ have higher $T_c$. Therefore, a much higher induced current is required to drive those regions, which constitute the percolation paths, to the normal state in VDP configuration.
$T_c$ of the percolation path corresponding to $R_8$ is smaller, and a lower current to drive it in the resistive state is required in two-probe configuration.

\section{Discussion\label{DISC}}
Results of the back gate experiments on our a-LTO/STO samples can be easily separated in three regions: i) negative gate voltages, ii) voltages between -20~V and +75~V, and iii) above +75~V. In the first region, transport is is governed by a one-band regime. Note that we do not observe an insulating state in the negative gate voltage range. This can be a sign of nonuniform conductivity. The behavior under voltage sweeps in the positive quadrant is another. We are apparently not able to fully trap the carriers and induce an insulating state as can occur in (crystalline) LAO/STO and LTO/STO interfaces \rev{ChunhaiPRL, BiscarasSREP}. Instead, we suggest that due to a significant non-uniformity of conducting properties, the trapping of electrons, which is seen in the hysteretic behavior, rearranges the current flow in the sample.
%Li et al. [\onlinecite{LiAS}] already showed in the non-stoichiometric LAO/STO samples that a metal state and superconductivity can be observed at much higher negative gate voltages than in stoichiometric LAO/STO. \textcolor{red}{Note - they actually say there is more dxyz, so why then one band ?} {\it Furthermore, they pinpoint a different subband distribution and have observed that \dxyz{} band is lying much lower, and the superconducting layer is thicker than in normal LAO/STO.}
%Since our layer is amorphous, our samples also have a different subband distribution when 2DES in optimized LAO/STO.

In the second regime the transport has changed to two band behavior. In this region, the MR exhibits the enhancement of out-of-plane and in-plane MR in agreement with previous works. Anisotropic in-plane MR has been reported in LAO/STO heterostructures\rev{BenShalomPRB, AriandoNCOM,WangPRB}. This behavior has been attributed to the magnetic ordering\rev{BenShalomPRB, WangPRB}.
Simultaneously, our observation of a bell shape of the in-plane MR at different gate voltage is similar to the results obtained by Diez et al. [\onlinecite{DiezPRL}]. They argued (see also Ref. [\onlinecite{BovenziPRB}]) that the decease in resistance, observed when the field is applied parallel to the plane and perpendicular to the current, can be described by a single particle Boltzmann equation. They showed that, when the second band is occupied, both interband scattering and spin-orbit coupling (SOC) are enhanced, which leads to the observed large negative in-plane MR. The MR is strongly modified in the gate region with the strongest SOC tunability, which would correspond to the region between 0~V  and 75~V in our data. However, we also see an unexpected enhancement in the geometry with current parallel to the field. We cannot exclude contribution of currents perpendicular to the field in this geometry, as mentioned in Sec. \ref{MR}, but another contribution may well arise from (spatial) mobility and carrier density fluctuations in our sample. In this region, $T_c$ and $I_c$ of superconducting state reach their maximum. \\

The high positive gate voltage range above 75~V is the range where the positive quasilinear MR develops which we believe is another signature of inhomogeneous transport in our films. In fact, such a crossover is observed in various different systems where spatial inhomogeneities can be invoked\rev{ParishNAT, KhouriPRL, HuPRB, KisslingerPRB, RamakrishnanPRB}. Generally, to observe the crossover at low fields requires relatively high mobilities. In our system these are available through high mobility carriers above the Lifshitz point.

Earlier, Ref. [\onlinecite{AriandoNCOM}] argued that the large positive MR supports an electronic phase separation scenario. However, there is a significant difference for our films compared to the ones studied in Ref. [\onlinecite{AriandoNCOM, WangPRB}]. Our system does not (for gate voltages of 0~V and above) exhibit an upturn of sheet resistance at low temperatures. Even more below 30 K, the MR for field-perpendicular-current is always negative. The main reason for this is that the results reported in Ref. [\onlinecite{AriandoNCOM}] were on  crystalline LAO/STO samples grown at the high pressure of $10^{-2}$ mbar \ox{}. Lower pressures leads to a decrease in the maximum magnetization according to results of Ref. \onlinecite{AriandoNCOM}, thus, making scenario of the phase separation between normal and magnetic region implausible as the main driving mechanism for the observed quasilinear MR.

At higher carrier densities (above 75 V), the in-plane MR showed a decrease, indicating an additional contribution which saturates in high fields. A connection between a non-trivial negative in-plane MR and a linear out-of-plane MR was actually observed in work on thin films of the Dirac semimetal Cd\Ss{3}As\Ss{2}\rev{SchumannPRB}, and in electron doped GaAs quantum wells\rev{JingNCOM}. In both cases, the macroscopic disorder is argued to be the origin of such behavior of MR. Additional support for this scenario in our samples is that the quasilinear MR develops in the region where high and low-mobility carriers have very similar carrier concentration as shown in Fig.\ref{Fig4}a, and even appear to cross.
So far, such crossing in STO-heterostructures has been only observed in experiments with top gate\cite{SanderPRL}. In Cd\Ss{3}As\Ss{2} an increase of negative MR was observed in the temperature range where two electron-type carriers have a crossover. However, in our case, a negative MR in current-perpendicular-field is also expected to arise from SOC effects and interband scattering. Spatial fluctuations in the conductivity can result in the current paths perpendicular to the magnetic field in in-plane geometry with the current parallel to the field\rev{SchumannPRB, HuPRB}. Together with the imperfection of the geometry used in the sample, it can lead to the non-trivial MR for this configuration.
Finally, also, the low temperature data point to the development of regions that do not become superconducting above 100 V and again indicate spatial fluctuations of conductivity. \\

Coming back to the superconductivity, extensive research already indicated the existence of inhomogeneous superconductivity in STO-based oxide heterostructures\rev{CapraraPRB2,BiscarasNMAT,DaptaryPRB,PrawiroatmodjoPRB,ThierschmannNCOM,HurandPRB}.
%One of the proposed model to describe phenomenology of the superconducting transition was Random Resistance Network model in Effective Mean Field Theory approach and provided insight on some properties of the~oxide interface such as temperature dependence\rev{CapraraPRB2} and $I(V)$ characteristics\rev{CapraraBKT}.
%In this model, large superconducting islands with different $T_{c}$ are randomly distributed over the sample.
As we discussed the behaviour of both $R(T)$ and $I(V)$ in our sample indicates the presence of strong spatial variations.
%In this case, Effective Mean Theory cannot be applied directly to experimental data\rev{CapraraPRB2} since it can fit only averaged curve.
%Through Random Resistance Network model still should be able to qualitatively reproduce results, but without knowledge of a real distribution of local $T_{c}$ it can be a challenging task.
The simple model we use to describe the inhomogeneous superconductor\rev{VaglioPRB} can describe some of the main features of the superconducting transition and critical current behaviour in our samples, although it is obviously too simple to be able to explain all the details of the real system, and in particular features arising due to a weak coupling between regions. \\

The final point to discuss is the possible origin of inhomogeneous electronic structure of the interface. This is the more important since it is often assumed that amorphous layers per se need not yield significantly different physics than crystalline layers. Previously, inhomogeneities in the conductance have been shown to arise from ferroelastic domains\rev{KaliskyNMAT,NoadPRB,FrenkelPRB,MaPRL,HonigNMAT}, which strongly affect superconducting properties\rev{NoadPRB,PaiPRL}. At the same time, as was mentioned, the quasilinear MR in our samples is much higher  than in the crystalline LAO/STO system, indicating an additional significant source of inhomogeneities. A prime candidate is  (oxygen) stoichiometry variations, most likely created during the growth. The amorphicity of the LTO layer itself may be an issue, but also the process of amorphization of LTO is not controlled in our samples, which can in particular be seen from the fact that RHEED oscillations were observed during growth. With respect to the amorphicity, it is instructive to note that also the deposition of amorphous LAO on STO led to a superconducting state which was described as a random array of Josephson-coupled superconducting domains\cite{PrawiroatmodjoPRB}.

\section{Conclusions}

We have grown and studied heterostructures of \LTO{}/\STO{}. In spite of clear two-dimensional growth, our samples were found to be amorphous, which may be due tot the absence of a capping layer. The samples showed the salient characteristics of the electron gas at oxide interfaces, in particular two-band behavior with normal values for the carrier concentrations and mobilities, as well as the existence of a Lifshitz point upon applying a gate voltage. The conductance was found to be high and inhomogeneous, signaled in particular by a large quasilinear MR and a percolative superconducting transition. By measuring in different configurations, both van der Pauw and two-probe, and using a simple model for a non-uniform superconductor\rev{VaglioPRB}, we were able explain prominent features of the superconducting transition in our sample. We propose that the non-uniformities arise from oxygen stoichiometry variations in our samples.

\begin{acknowledgments}
N.L. and J.A. gratefully acknowledge the financial support of the research program DESCO, which is financed by the Netherlands Organisation for Scientific Research (NWO). The authors thank J. Jobst, S. Smink, K. Lahabi and G. Koster for useful discussion.
\end{acknowledgments}

\bibliography{LTO_IS_aug20}

\end{document}

\newcommand{\Ss}{\textsubscript}
\newcommand{\Us}{\textsuperscript}
\newcommand{\STO}{SrTiO\Ss{3}}
\newcommand{\GTO}{GdTiO\Ss{3}}
\newcommand{\LAO}{LaAlO\Ss{3}}
\newcommand{\LTO}{LaTiO\Ss{3}}
\newcommand{\RTO}{ReTiO\Ss{3}}
\newcommand{\NGO}{NdGaO\Ss{3}}
\newcommand{\LCO}{LaCrO\Ss{3}}
\newcommand{\PyroCl}{Re\Ss{2}Ti\Ss{2}O\Ss{7}}
\newcommand{\ETO}{EuTiO\Ss{3}}
\newcommand{\ox}{O\Ss{2}}
\newcommand{\tit}{Ti\Us{3+}}
\newcommand{\tif}{Ti\Us{4+}}
\newcommand{\DC}{\degree C}
\newcommand{\mx}{\times}
\newcommand{\rev}{\cite}
\newcommand{\RAN}{$R^{xy}_{AN}$}
\newcommand{\dxy}{3d\Ss{xy}}
\newcommand{\dxyz}{3d\Ss{xz/yz}}
\newcommand{\dxyzo}{d\Ss{xz/yz}}
\newcommand{\dxyo}{d\Ss{xy}}
\renewcommand\thefigure{S\arabic{figure}}
\renewcommand\theequation{S\arabic{equation}}
%\preprint{}

\title{Supplementary information}


\author{N. Lebedev}
%\thanks{}
\affiliation{Huygens - Kamerlingh Onnes Laboratory, Leiden University, P.O. Box 9504, 2300 RA Leiden, The Netherlands}
\author{M. Stehno}
\affiliation{Physikalisches Institut (EP 3), Universität Würzburg, Am Hubland 97074 Würzburg, Germany}
\author{A. Rana}
\affiliation{School of Engineering and Technology, BML Munjal University (Hero Group), Gurgaon, India - 122413}
\author{N. Gauquelin}
\author{J. Verbeeck}
\affiliation{Electron Microscopy for Materials Science, University of Antwerp, Campus Groenenborger Groenenborgerlaan 171, 2020 Antwerpen, Belgium}
\author{A. Brinkman}
\affiliation{MESA+ Institute for Nanotechnology, University of Twente, P.O. Box 217, 7500 AE Enschede, The Netherlands}
\author{J. Aarts}
\affiliation{Kamerlingh Onnes Laboratory, Leiden University, P.O. Box 9504, 2300 RA Leiden, The Netherlands}

\date{\today}

%\pacs{}

\maketitle


\section{Analysis of carrier concentrations and mobilities} \label{Carr}

To extract carrier concentration and mobility, we used the following equation in the two-band region:
\begin{equation}\label{eq1}
R_{xy}(B)= - \frac{B}{e} \frac{n_1\mu_1^2+n_2\mu_2^2+(n_1+n_2)\mu_1^2\mu_2^2B^2}{(n_1\mu_1+n_2\mu_2)^2+(n_1+n_2)^2\mu_1^2\mu_2^2B^2},\\
\end{equation}
where $e$ is the elementary charge, $n_i$ and $\mu_i$ are carrier concentration and mobility of $i$ type carriers correspondingly. For fitting the constraint $R_S(0)=1/e(n_1\mu_1+n_2\mu_2)$ was used.
In the one-band region, we used high field Hall resistance value and zero field sheet resistance.

\section{Magnetotransport properties}\label{GD}
\bigskip
\begin{figure}[h]
\includegraphics[scale=0.5]{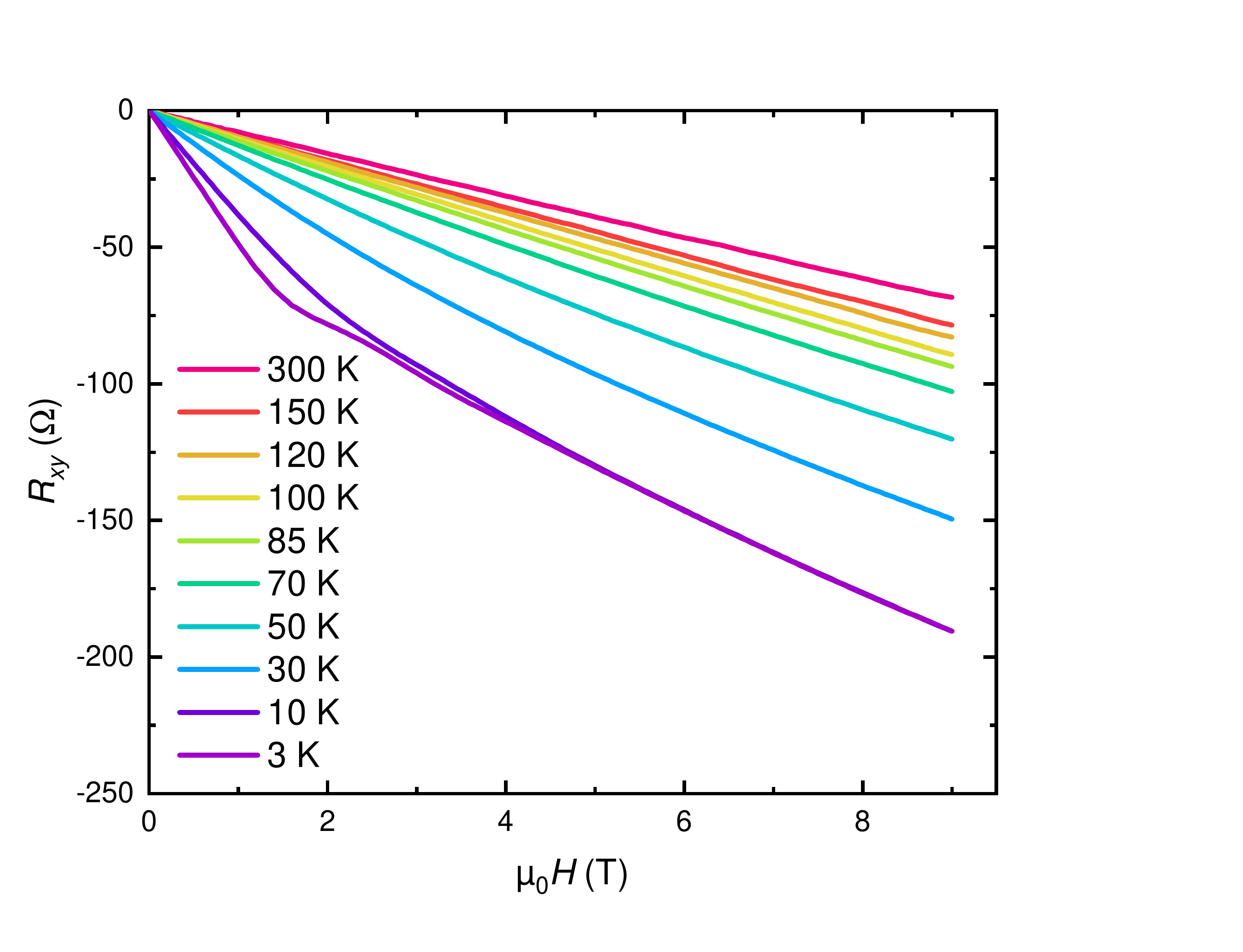}
\caption{Hall resistance at different temperature.\label{Fig1}}
\end{figure}
\begin{figure}[h]
\includegraphics[scale=0.35]{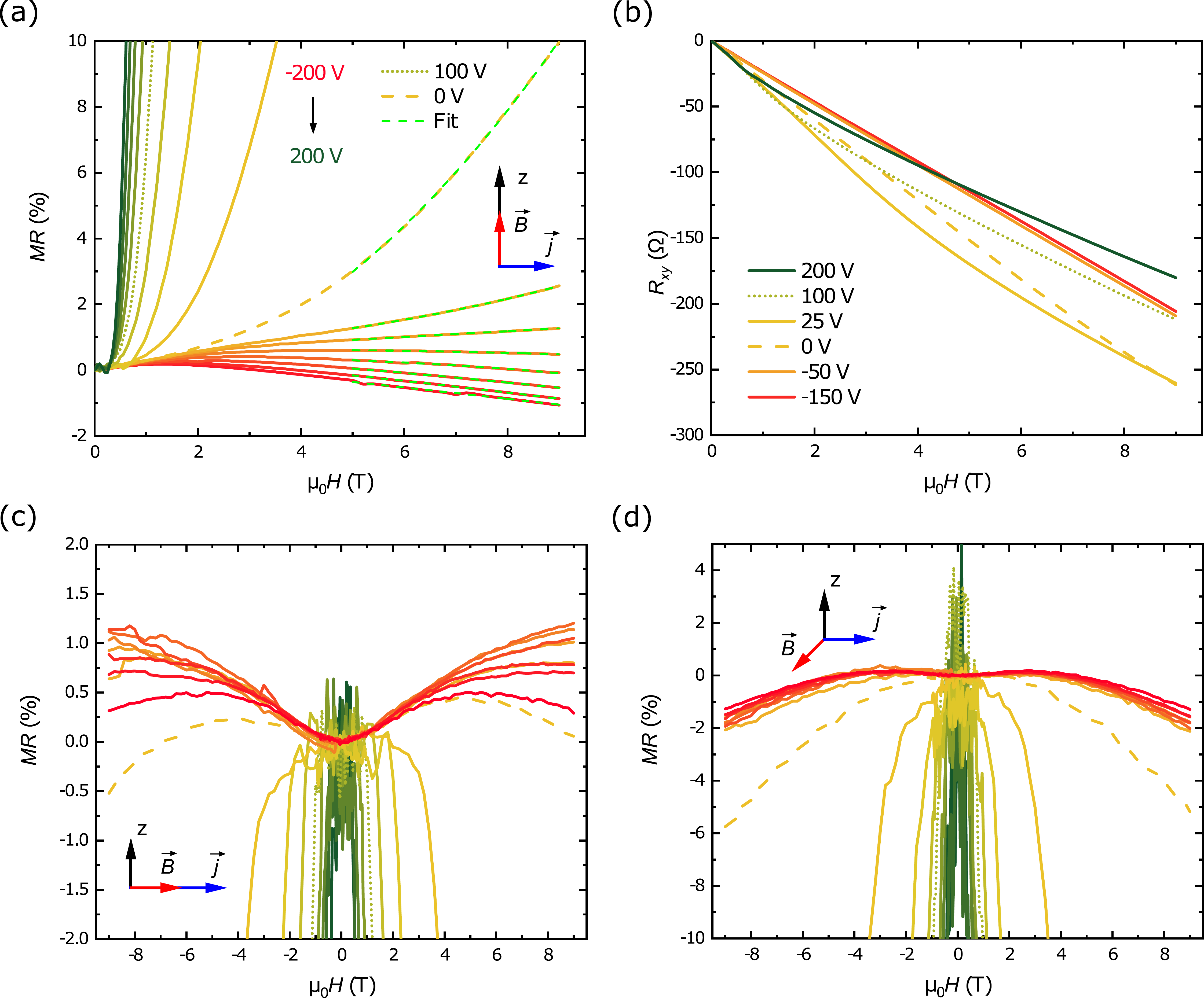}
\caption{(a) A zoomed-in part of the MR at $3$ K with the magnetic field perpendicular to the sample plane. (b) Hall resistance at $3$ K for selected gate voltage values. (c) A zoomed-in part of the MR at $3$ K with the magnetic field in-plane and parallel to the current direction and (d) in-plane perpendicular to the current direction.\label{Fig2}}
\end{figure}
\clearpage
\section{Superconducting properties\label{SC}}

\begin{figure}[h]
\includegraphics[scale=0.25]{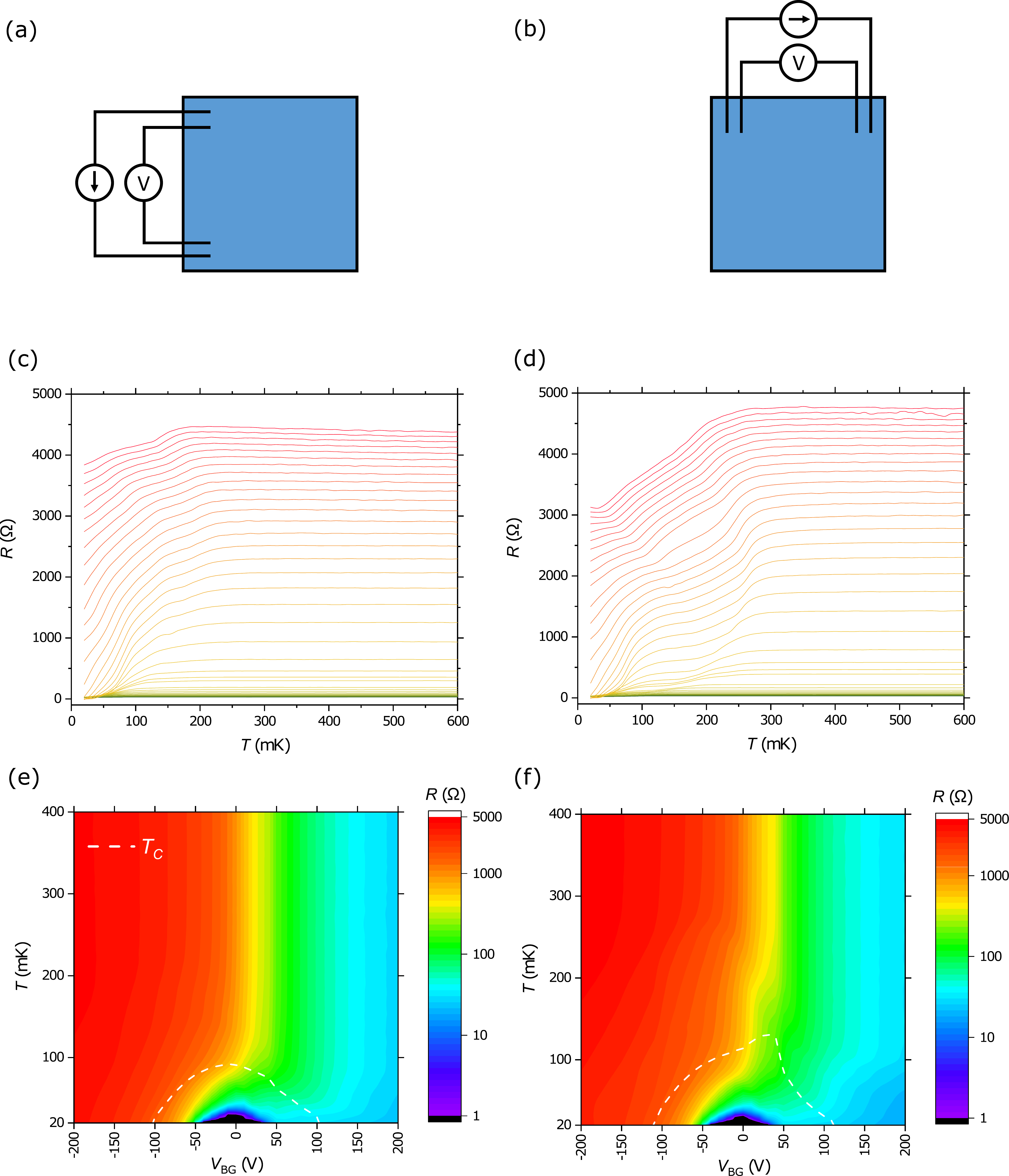}
\caption{Behavior of the resistance as function of temperature T and back gate voltage V$_{BG}$ for two two-probe contact geometries called (a) 'vertical' and (b) 'horizontal'. $R(T)$ curves at different gate voltage with step of $10$ V for (c) vertical and (d) horizontal configurations. The same data visualized in colour map form for (e) vertical and (f) horizontal configurations.\label{Fig3}}
\end{figure}

\clearpage

\begin{figure}[h]
\includegraphics[scale=0.17]{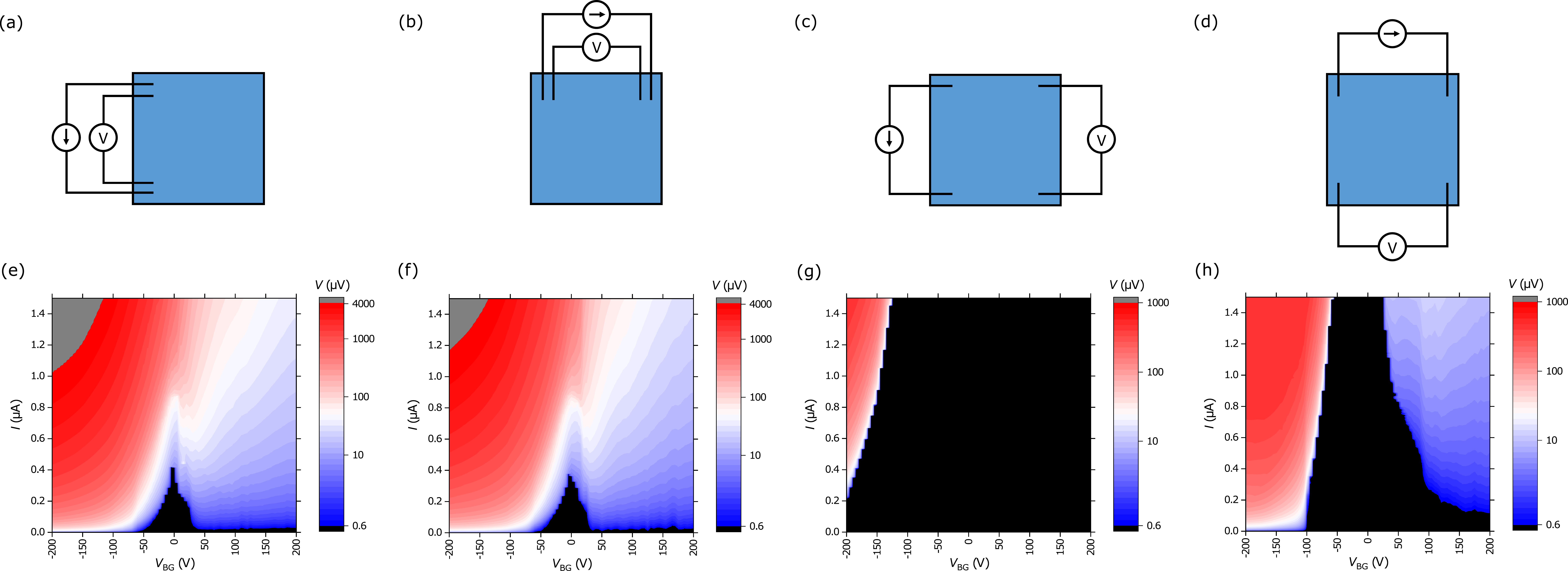}
\caption{(a)-(d) Schematics of measurements geometry. (e)-(h) $I(V)$ curves at the different gate voltages at 20 mK.\label{Fig4}}
\end{figure}

\section{Circuit analysis}

\begin{figure}
\includegraphics[scale=0.75]{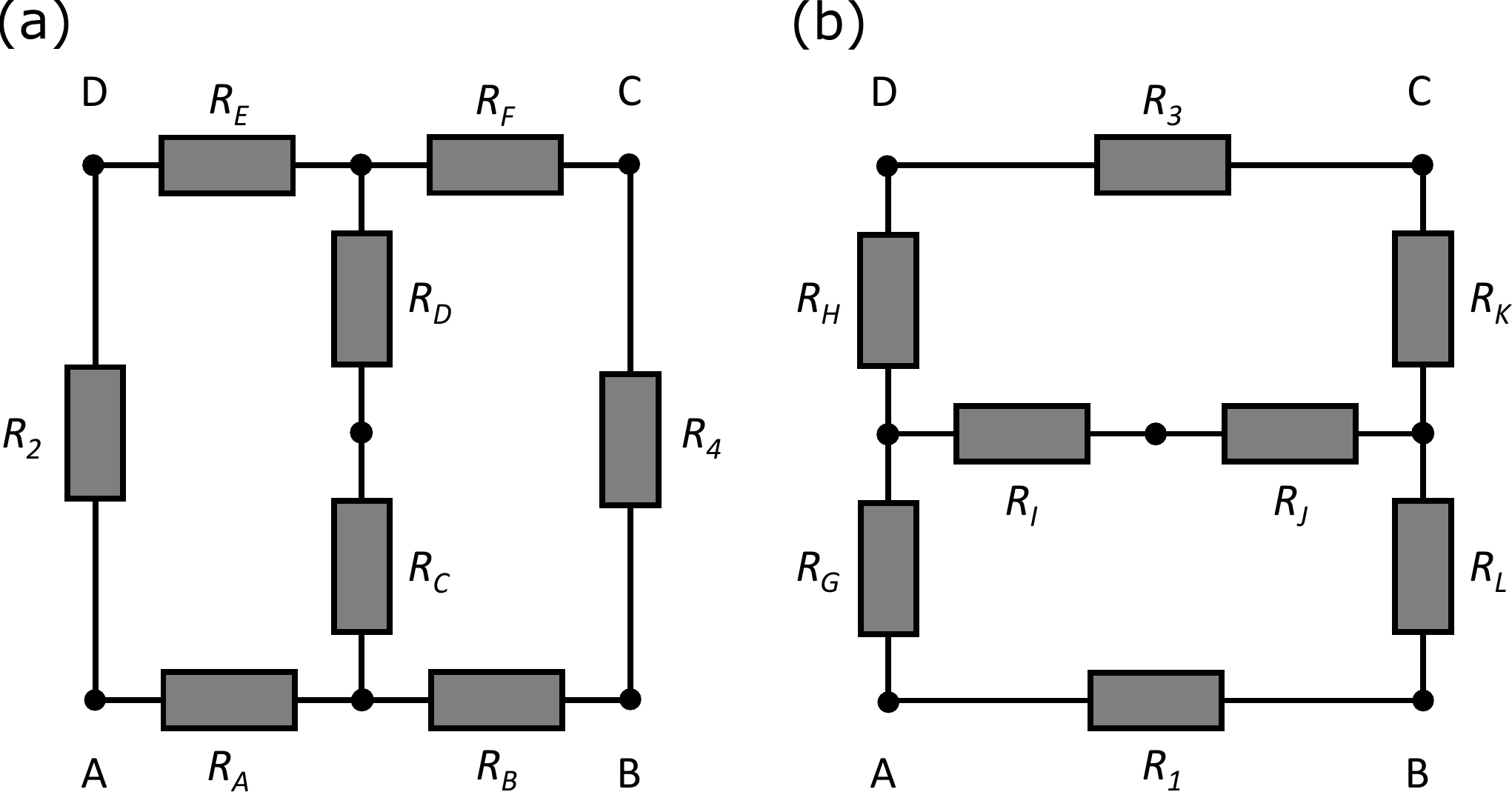}
\caption{(a) An alternative representation of the model circuit from the main text for (a) vertical and (b) horizontal configurations.\label{Fig5}}
\end{figure}

The measured resistances can be expressed through a set of resistances of the model in the following way.
First, using a Y-$\Delta$ network transform, it is possible to simplify the model circuit into two simple ones for the vertical and horizontal direction. The circuits obtained in this way are shown in Fig. \ref{Fig5}.
The newly introduced set of resistances are expressed as:
\begin{equation}\label{eq2}
R_{A}=\frac{R_{1}R_{5}}{R_{1}+R_{5}+R_{6}},
R_{B}=\frac{R_{1}R_{6}}{R_{1}+R_{5}+R_{6}},
R_{C}=\frac{R_{5}R_{6}}{R_{1}+R_{5}+R_{6}},\\
\end{equation}
\begin{equation}\label{eq3}
R_{D}=\frac{R_{7}R_{8}}{R_{3}+R_{7}+R_{8}},
R_{E}=\frac{R_{3}R_{8}}{R_{3}+R_{7}+R_{8}},
R_{F}=\frac{R_{3}R_{7}}{R_{3}+R_{7}+R_{8}},\\
\end{equation}
\begin{equation}\label{eq4}
R_{G}=\frac{R_{4}R_{5}}{R_{4}+R_{5}+R_{8}},
R_{H}=\frac{R_{4}R_{8}}{R_{4}+R_{5}+R_{8}},
R_{I}=\frac{R_{5}R_{8}}{R_{4}+R_{5}+R_{8}},\\
\end{equation}
\begin{equation}\label{eq5}
R_{J}=\frac{R_{6}R_{7}}{R_{2}+R_{6}+R_{7}},
R_{K}=\frac{R_{2}R_{7}}{R_{2}+R_{6}+R_{7}},
R_{L}=\frac{R_{2}R_{7}}{R_{2}+R_{6}+R_{7}},\\
\end{equation}
These circuits can be solved by the mesh current method.
After some lengthy but straightforward arithmetic calculations, the resistances in the different measurement configurations can be expressed as follow:
 \begin{equation}\label{eq6}
  R_{VDP}^{hor}=\frac{\pi}{\ln  2}R_{CD,BA}=\frac{\pi}{\ln  2}\frac{V_{BA}}{I_{CD}}=\frac{\pi}{\ln  2}\frac{R_{1}R_{3}(R_{2}R_{5}R_{8}+R_{4}R_{6}R_{7}+Y)}{Z},\\
 \end{equation}
 \begin{equation}\label{eq7}
 R_{VDP}^{ver}=\frac{\pi}{\ln  2}R_{AD,CB}=\frac{\pi}{\ln  2}\frac{V_{CB}}{I_{AD}}= \frac{\pi}{\ln  2}\frac{R_{2}R_{4}(R_{1}R_{7}R_{8}+R_{3}R_{5}R_{6}+Y)}{Z},\\
 \end{equation}
 \begin{equation}\label{eq8}
 \begin{split}
 R_{2 probe}^{hor}=R_{CD,CD}=\frac{V_{CD}}{I_{CD}}= \frac{R_{3}(U+(R_{1}+R_{2}+R_{4})Y+R_{4}W+X)}{Z}
 \end{split}
 \end{equation}
 \begin{equation}\label{eq9}
 \begin{split}
R_{2 probe}^{ver}=R_{AD,AD}=\frac{V_{AD}}{I_{AD}}= \frac{R_{4}(V+(R_{1}+R_{2}+R_{3})Y+R_{3}W+X)}{Z},\\
 \end{split}
 \end{equation}
 \begin{equation}\label{eq10}
 \begin{split}
  W=R_{8}(R_{2}(R_{1}+R_{5}+R_{6})+R_{1}(R_{6}+R_{7})),\\
  X=R_{1}R_{2}(R_{5}R_{7}+R_{8}(R_{5}+R_{7})),\\
  Y=R_{5}R_{7}(R_{6}+R_{8})+R_{6}R_{8}(R_{5}+R_{7}),\\
  U=R_{4}R_{7}(R_{2}(R_{1}+R_{5}+R_{6})+R_{1}R_{6}),\\
  V=R_{3}R_{5}(R_{1}(R_{2}+R_{6}+R_{7})+R_{2}R_{6}),\\
  Z=(R_{1}+R_{2}+R_{3}+R_{4})Y+X+(R_{3}+R_{4})W+U+V+\\
  R_{3}R_{4}((R_{1}+R_{5}+R_{6})(R_{2}+R_{7})+R_{6}(R_{1}+R_{5})),
 \end{split}
 \end{equation}
where $V_{ij}=V_{i}-V_{j}$, and $I_{ij}$ is the current injected in contact $i$ and taken out from contact $j$.
$R_i$ is assumed to follow a simple stepwise function for an inhomogeneous phase transition\rev{VovkJMS}:
\begin{equation}\label{eq11}
R_i=\frac{R_{norm_i}}{(1+e^{(T_{c_i}-T)/w_i)})},\\
\end{equation}
where $R_{norm_i}$ is resistance in normal state, $T_{c_i}$ is the critical temperature, which corresponds to the resistance with $R_{norm_i}/2$ and $w_i$ determines the broadening of the superconducting transition.
The result of the fitting is present in Table \ref{Tab1}. Note that parameters have been limited according to the picture described in the main text, and some of them have been fixed in such a way that the resistance stays constant in the measured range.
\begin{table}[h]
\caption{\label{Tab1}The parameters obtained from fits of $R_{VDP}^{horz}$, $R_{VDP}^{vert}$, $R_{2 probe}^{horz}$ and $R_{2 probe}^{vert}$ measured at $0$~V with Eq. \ref{eq6}-\ref{eq10}.}
\begin{ruledtabular}
\begin{tabular}{ccccc}
$R_i$ & $R_{norm}$ ($\Omega$)& $T_i$ (K) & $w_i$ (K) \\ \hline
$R_1$\footnote{$T_i$ and $w_i$ are fixed.} & $1453.3 \pm 310.7$ & $0.001$ & $0.0001$\\
$R_2$ & $2644.8 \pm 813.3$ & $0.247 \pm 0.005$ & $0.0178 \pm 0.0002$ & \\
$R_3$\footnotemark[1] & $3626.8 \pm 623.2$ & $0.001$ & $0.0001$ \\
$R_4$\footnotemark[1] & $4289.0 \pm 1377.5$ & $0.001$ & $0.0001$\\
$R_5$ & $1098.4 \pm 360.9$ & $0.13 \pm 0.005$ & $0.019 \pm 0.0003$\\
$R_6$\footnotemark[1] & $277.3 \pm 7.1$ & $0.001$ & $0.0001$\\
$R_7$ & $1569.3 \pm 497.0$ & $0.14 \pm 0.003$ & $0.014 \pm 0.0003$\\
$R_8$ & $608.4 \pm 16.8$ & $0.08 \pm 0.0004$ & $0.01 \pm 0.0001$\\
\end{tabular}
\end{ruledtabular}
\end{table}

\bibliography{LTO_IS_aug20}